\documentclass[conference,10pt]{IEEEtran}

\IEEEoverridecommandlockouts
 \makeatletter
 \def\endthebibliography{%
 	\def\@noitemerr{\@latex@warning{Empty `thebibliography' environment}}%
 	\endlist
 }
\usepackage{cite}
\usepackage{amsmath,amssymb,amsfonts}
\usepackage{algorithmic}
\usepackage{algorithm}
\usepackage{graphicx}
\usepackage{textcomp}
\usepackage{bm}
\usepackage{amsmath,amsfonts,amssymb,amsbsy,nccmath} \usepackage{amsthm}

\newcommand{\bmA}{\mathbf A}

\newcommand{\bmI}{\mathbf I}
\newcommand{\bmT}{\mathbf T}
\newcommand{\bmQ}{\mathbf Q}
\newcommand{\bmB}{\mathbf B}

\newcommand{\bmE}{\mathbf E}

\newcommand{\bmH}{\mathbf H}
\newcommand{\bmF}{\mathbf F}
\newcommand{\bmS}{\mathbf S}
\newcommand{\bms}{\mathbf s}

\newcommand{\bmV}{\mathbf V}
\newcommand{\bmU}{\mathbf U}

\newcommand{\bmK}{\mathbf K}
\newcommand{\bmy}{\mathbf y}
\newcommand{\bmZ}{\mathbf Z}
\newcommand{\bmJ}{\mathbf J}
\newcommand{\bmR}{\mathbf R}
\newcommand{\bmX}{\mathbf X}

\newcommand{\bmW}{\mathbf W}

\newcommand{\bmq}{\mathbf q}

\newcommand{\bmx}{\mathbf x}
\newcommand{\bmn}{\mathbf n}

\newcommand{\bme}{\mathbf e}

 \usepackage[left=1.62cm,right=1.62cm,top=1.9cm]{geometry} 
 
\usepackage{xcolor}

\begin{document}
\title{Robust Beamforming for IRS Aided MIMO Full Duplex Systems}
 
\author{Chandan~Kumar~Sheemar,
   \;Jorge Querol, Sourabh Solanki, 
   Sumit Kumar, and 
Symeon Chatzinotas \\
SnT, University of Luxembourg, emails: \{name.surname\}@uni.lu
}
\maketitle

\vspace{-5mm}
\begin{abstract}
In this paper, a novel robust beamforming for an intelligent reflecting surface (IRS) assisted FD system is presented. Since perfect channel state information (CSI) is often challenging to acquire in practice, we consider the case of imperfect CSI and adopt a statistically robust beamforming approach to maximize the ergodic weighted sum rate (WSR). We also analyze the achievable WSR of an IRS-assisted FD with imperfect CSI, for which the lower and the upper bounds are derived.
The ergodic WSR maximization problem is tackled based on the expected Weighted Minimum Mean Squared Error (WMMSE), which is guaranteed to converge to a local optimum. The effectiveness of the proposed design is investigated with extensive simulation results. It is shown that our robust design achieves significant performance gain compared to the naive beamforming approaches and considerably outperforms the robust Half-Duplex (HD) system.
\end{abstract}

 
\IEEEpeerreviewmaketitle

\vspace{-2mm}
\section{Introduction}
\IEEEPARstart{F}{ull Duplex} (FD) is a promising wireless transmission technology offering simultaneous transmission and reception in the same frequency band, which theoretically doubles the spectral efficiency \cite{sheemar2022practical,sheemar2021beamforming_5}.
 Beyond spectral efficiency, FD can be beneficial to improve security, enable advanced joint communication and sensing, and reduce end-to-end delays. However, FD systems suffer from Self-Interference (SI), which could be $90-110~$dB higher than the received signal of interest, and it is a major challenge for achieving an ideal FD operation. Advanced SI Cancellation (SIC) techniques are pivotal to mitigate the SI power close to the noise floor and enable correct reception of the received signal.
In parallel to FD, another emerging technology is the Intelligent Reflecting Surfaces (IRSs), which can provide a smart and reconfigurable wireless environment \cite{sharma2020intelligent,sheemar2021hybrid_1,sheemar2022hybrid_2}. 

 
The IRS-aided FD (IRS-FD) systems hold great promise in supporting the forthcoming traffic demands while being highly spectrally and energy efficient \cite{sheemar2023full_3,sheemar2022near_4,sheemar2020receiver_7}. In \cite{sharma2020intelligent}, the authors investigated the performance of the IRS-FD system by analyzing the outage and error probabilities. In \cite{abdullah2020optimization}, the authors present a novel beamforming design for an FD relay assisted with one IRS to maximize the minimum achievable rate with the max-min optimization. 
In \cite{ge2021robust}, the authors studied the IRS-FD systems to improve the worst-case achievable security rate under imperfect channel state information (CSI).
Furthermore, a deep neural network is also developed to reduce the computational burden of the proposed beamforming solution. Finally, in \cite{saeidi2021weighted}, the authors presented a 
joint beamforming design for weighted sum rate (WSR) maximization in a single-cell multiple-input single-output (MISO) IRS-FD system. Note that the existing literature only provides the secrecy rate expression for IRS-FD systems under imperfect Channel State Information (CSI). However, there is currently no analytical expression available for the ergodic weighted sum rate (WSR) or a novel robust beamforming design considering imperfect CSI in this context.

Motivated by the aforementioned considerations, we address the challenge of maximizing the ergodic WSR for IRS-FD systems with multi-antenna users under imperfect CSI. Our objective is to enhance both the uplink (UL) and downlink (DL) sum rates by jointly optimizing the digital beamformers and the IRS phase response while accounting for imperfect CSI. It is important to note that incorporating CSI errors in the beamforming designs, with multi-antenna users, makes the problem of MIMO beamforming for IRS-FD systems highly complex. The CSI errors are modelled by using the Gaussian-Kronecker model and then an analytical approximation for the ergodic achievable WSR using this model is derived. Subsequently, we formulate the problem of maximizing the ergodic WSR, considering average sum-power and unit-modulus constraints for the IRS phase response under imperfect CSI. To solve this problem, we leverage the connection between the ergodic WSR and the expected weighted minimum mean squared error (EWMMSE), where the expectation is taken with respect to the CSI errors. Simulation results validate the accuracy of the analytically derived ergodic WSR expression. It is further shown that our proposed approach significantly outperforms the naive schemes in the presence of imperfect CSI.\footnote{\emph{Notations:}  Boldface lower and upper case characters denote vectors and matrices, respectively. $\mathbb{E}\{\cdot\}$, $\mbox{Tr}\{\cdot\}$, $\bmI$, $\otimes$, $\widetilde{\bmX}$  denote expectation, trace, identity matrix, Kronecker product, and transmit covariance matrix, respectively. Finally, $\mbox{diag}(\bmx)$ and denote a diagonal matrix with vector $\bmx$ on its main diagonal and  $\bmI$ denotes identity matrix.}

\section{System Model} \label{sistema}
Let $j$ and $k$ denote the multi-antenna DL and UL user served by the MIMO FD base station (BS), respectively, and
let $M_{k}$ and $N_{j}$ denote their number of transmit and receive antennas, respectively.
The FD BS is assumed to have  $M_0$ transmit and $N_0$ receive antennas. We consider a multi-stream approach, and the number of data streams for the UL user $k$ and DL user $j$ is denoted as $u_k$ and $v_j$, respectively. Let $\bmU_k \in \mathbb{C}^{M_k \times u_k}$ and $\bmV_j \in \mathbb{C}^{M_t \times v_j}$ denote the precoders for white unitary-variance data streams $\bms_{k}\in \mathbb{C}^{u_k \times 1}$ and $\bms_{j} \in \mathbb{C}^{v_j \times 1}$, respectively. We assume that the considered FD system is aided with one IRS of size $R \times C$.
Let $\bm{\theta} = [e^{i \theta_1},....,e^{i \theta_{RC}}]$ denote the vector containing the phase-shift response of its $RC$ elements, and let $\mathbf{\Theta} = \mbox{diag}(\bm{\theta})$ denote a diagonal matrix containing $\bm{\theta}$ on its main diagonal. Let $\bmn_0$ and 
$\bmn_j$ denote the noise vectors at the FD BS and DL user $j$, respectively, which are modelled as $\bmn_0 = \mathcal{CN}(\bm0,\sigma_0^2 \bmI)$ $ \bmn_j = \mathcal{CN}(\bm0,\sigma_j^2 \bmI)$,
where  $\sigma_0^2$ and $\sigma_j^2$ denote the noise variances at the FD node and the DL user $j$, respectively. The channel responses from the UL user $k$ to the BS and from the BS to the DL user $j$ are denoted with  $\bmH_k \in \mathbb{C}^{N_0 \times M_k}$ and $\bmH_j \in \mathbb{C}^{N_j \times M_0}$, respectively. Let $\bmH_0\in  \mathbb{C}^{N_0 \times M_0}$ and $\bmH_{j,k} \in  \mathbb{C}^{N_j \times M_k}$ denote the SI channel response for the FD BS and cross-interference channel response between the UL user $k$ and the DL user $j$, respectively. The channel responses from the transmit antenna array of the FD BS to the IRS and from the IRS to the receive antenna array of the FD BS are denoted with $\bmH_{\theta,0} \in \mathbb{C}^{R C \times M_0}$
and $\bmH_{0,\theta} \in \mathbb{C}^{N_0 \times R C}$, respectively. Let $\bmH_{j,\theta} \in \mathbb{C}^{N_j \times RC}$ and $\bmH_{\theta,k} \in \mathbb{C}^{RC \times M_k}$ denote the channel responses from the IRS to the DL user $j$ and from the UL user $k$ to the IRS, respectively.

 \vspace{-1mm}

\subsection{Imperfect CSI Modelling}
Let $\Delta\bmH_k, \Delta\bmH_j, \Delta\bmH_0, \Delta\bmH_{j,k}, \Delta\bmH_{\theta,0}, \Delta\bmH_{0,\theta}, \Delta\bmH_{j,\theta}, \Delta\bmH_{\theta,k}$ denote the estimation errors for the channel responses $\bmH_k, \bmH_j, \bmH_0, \bmH_{j,k}, \bmH_{\theta,0}, \bmH_{0,\theta}, \bmH_{j,\theta}, \bmH_{\theta,k}$, respectively. The available CSI matrices can be written as a sum of the channel estimates and CSI errors as
\begin{equation} \label{channel_decomp}
    \begin{aligned}
     & \bmH_k  =  \hat{\bmH}_k + \Delta \bmH_k, \quad  
      \quad \quad \quad \quad \quad \quad\; \bmH_j  = \hat{\bmH}_j + \Delta \bmH_j, \\
     & \bmH_0  = \hat{\bmH}_0 + \Delta \bmH_0, \quad \quad \quad \quad \quad  \quad \quad\;\;
      \bmH_{j,k}= \hat{\bmH}_{j,k} + \Delta \bmH_{j,k}, \\
      &\bmH_{\theta,0} = \hat{\bmH}_{\theta,0} + \Delta \bmH_{\theta,0}, \quad \quad \quad \quad \quad \;
     \bmH_{0,\theta} = \hat{\bmH}_{0,\theta} + \Delta\bmH_{0,\theta}\\
    &  \bmH_{j,\theta} = \hat{\bmH}_{j,\theta} + \Delta\bmH_{j,\theta}, \quad \quad \quad \quad \quad \;
      \bmH_{\theta,k} = \hat{\bmH}_{\theta,k} + \Delta\bmH_{\theta,k},\\
    \end{aligned}      
\end{equation}
where the channel matrices of the form $\hat{\bmX}$ denote the channel estimates. To model the estimation errors, we adopt the Gaussian Kronecker model \cite{rong2011robust}, which dictates that

\begin{equation} \label{CE_modelling}
    \begin{aligned}
       &  \Delta \bmH_k = \mathcal{CN}(\bm0,\bmJ_k  \otimes \bmK_k ), \quad  
        \;\; \Delta \bmH_{\theta,0} = \mathcal{CN}(\bm0,\bmJ_{\theta,0} \otimes \bmK_{\theta,0} ), \\
    &  \Delta \bmH_0 = \mathcal{CN}(\bm0,\bmJ_0  \otimes \bmK_0 ), \quad   \quad
      \Delta \bmH_{j,k} = \mathcal{CN}(\bm0,\bmJ_{j,k} \otimes \bmK_{j,k}), \\
      & \Delta \bmH_j = \mathcal{CN}(\bm0,\bmJ_j  \otimes \bmK_j), \quad\; \quad  \quad
     \Delta\bmH_{0,\theta} = \mathcal{CN}(\bm0,\bmJ_0  \otimes \bmK_{\theta}), \\
    &  \Delta\bmH_{j,\theta} =\mathcal{CN}(\bm0,\bmJ_{j,\theta}  \otimes \bmK_{j,\theta} ), \;\;  \Delta\bmH_{\theta,k} \mathcal{CN}(\bm0,\bmJ_{\theta,k}  \otimes \bmK_{\theta,k} ),\\
    \end{aligned}          
\end{equation}
where the matrices $\bmJ$ and $\bmK$ are the covariance matrices seen from the transmitter and receiver for each link, respectively. The estimation errors are uncorrelated with the estimated channel matrices, and hence we can write 
\begin{equation} \label{Channel_modelling_err}
   \begin{aligned}
      &  \bmH_k = \mathcal{CN}(\hat{\bmH}_k,\bmJ_k  \otimes \bmK_k), \;\;\;
        \bmH_{\theta,0} = \mathcal{CN}(\hat{\bmH}_{\theta,0},\bmJ_{\theta,0} \otimes \bmK_{\theta,0}), \\  
     & \bmH_0 = \mathcal{CN}(\hat{\bmH}_0,\bmJ_0  \otimes \bmK_0), \;\;\;
     \bmH_{j,k} = \mathcal{CN}(\hat{\bmH}_{j,k},\bmJ_{j,k} \otimes \bmK_{j,k}), \\
      & \bmH_j = \mathcal{CN}(\hat{\bmH}_j,\bmJ_j  \otimes \bmK_j), \;\;\;\;\bmH_{\theta,k}= \mathcal{CN}(\hat{\bmH}_{\theta,k},\bmJ_{\theta,k}  \otimes \bmK_{\theta,k})\\
 & \bmH_{j,\theta} \hspace{-0.4mm}=\hspace{-0.4mm}\mathcal{CN}(\hat{\bmH}_{j,\theta},\bmJ_{j,\theta}  \otimes \bmK_{j,\theta}) \bmH_{0,\theta} \hspace{-0.2mm} =  \hspace{-0.4mm} \mathcal{CN}(\hat{\bmH}_{0,\theta},\bmJ_{0,\theta} \otimes \bmK_{0,\theta})  
    \end{aligned}   
\end{equation}

Let $\overline{\bmH}_{0,k}, \overline{\bmH}_{0,0}, \overline{\bmH}_{j},\overline{\bmH}_{j,k}$ denote the effective channel responses affected by the estimation errors defined as  

\begin{subequations} \label{eff_channel}
\begin{equation}
   \overline{\bmH}_{k} = (\hat{\bmH}_k + \Delta \bmH_k ) +  (\hat{\bmH}_{0,\theta}  + \Delta \bmH_{0,\theta} ) \mathbf{\Theta} (\hat{\bmH}_{\theta,k} + \Delta\bmH_{\theta,k}) 
\end{equation} 
\begin{equation}
  \overline{\bmH}_{0} =  (\hat{\bmH}_0 + \Delta \bmH_0) + (\hat{\bmH}_{0,\theta} + \Delta \bmH_{0,\theta} ) \mathbf{\Theta} (\hat{\bmH}_{\theta,0} + \Delta\bmH_{\theta,0}) 
\end{equation}
\begin{equation}   
\overline{\bmH}_{j}=(\hat{\bmH}_j + \Delta \bmH_j) + (\hat{\bmH}_{j,\theta}  + \Delta \bmH_{j,\theta}) \mathbf{\Theta} (\hat{\bmH}_{\theta,0}  + \Delta\bmH_{\theta,0}) 
\end{equation}
\begin{equation}
   \overline{\bmH}_{j,k} = (\hat{\bmH}_{j,k} + \Delta \bmH_{j,k})  +  (\hat{\bmH}_{j,\theta} +\Delta\bmH_{j,\theta}) \mathbf{\Theta} (\hat{\bmH}_{\theta,k} + \Delta  \bmH_{\theta,k} ) 
\end{equation}
\end{subequations}
Let $\bmy_k$ and $\bmy_j$ denote the signals received by the FD BS from UL user $k$ and by the DL user $j$, respectively, given as \vspace{-2.2mm}
\begin{subequations} \label{Rx_side_CE}
    \begin{equation}  
       \bmy_k = \overline{\bmH}_k \bmU_k  \bms_{k} +   \overline{\bmH}_0 \bmV_j \bms_{j} + \bmn_0, 
\end{equation}
\begin{equation}
         \bmy_j = \overline{\bmH}_j  \bmV_j  \bms_j  + \overline{\bmH}_{j,k}  \bmU_k  \bms_{k} + \bmn_j.
\end{equation}  \vspace{-4mm}
\end{subequations}
\subsection{Problem Formulation}
We aim to maximize the ergodic WSR of the IRS-FD system under imperfect CSI given the CSI error statistics and the channel estimates. Let $\mathcal{R} = \mathcal{R}_k + \mathcal{R}_j$ denote the WSR of the system in the case of perfect CSI, with $\mathcal{R}_k$ and $\mathcal{R}_j$ denoting the weighted rate of the users $k$ and $j$, respectively. The ergodic WSR of the system, i.e., the average WSR with respect to the CSI errors, can be written as $\mathbb{E}_{\bmH|\hat{\bmH}}[\mathcal{R}]$. However, such a problem is infeasible to be solved and can be tackled by applying Jensen's inequality, which allows moving the expectation operator insider as  $\mathbb{E}_{\bmH|\hat{\bmH}}[\mathcal{R}] \geq \mathcal{R}(\mathbb{E}_{\bmH|\hat{\bmH}})$ \cite{negro2012sum}. The optimization problem for ergodic WSR $\mathcal{R}(\mathbb{E}_{\bmH|\hat{\bmH}})$,  under the sum-power and the unit-modulus constraint can be written as
 
\begin{subequations}\label{WSR_problem}
\begin{equation}
\underset{\substack{\bmV_j,\bmU_k},\mathbf{\Theta}}{\max} \quad  \mathcal{R}_k(\mathbb{E}_{\bmH|\hat{\bmH}}) + \mathcal{R}_j(\mathbb{E}_{\bmH|\hat{\bmH}}) \vspace{-4mm}
\end{equation} \label{WSR} 
\begin{equation} \vspace{-1mm}
\text{s.t.}  \quad\mbox{Tr}( \bmU_k \bmU_k^H ) 	\preceq \alpha_k, \; \&  \; \mbox{Tr} ( \bmV_j  \bmV_j^H ) \leq  \alpha_0, \label{c2}
\end{equation}
\begin{equation} \label{c3} \vspace{-1mm}
 \quad 
  \quad   \quad  |\bm{\theta}(i)|=1,  \quad \forall i,
\end{equation}
\end{subequations}
where  $\alpha_0$ and $\alpha_k$ denote the sum-power constraint at the BS and UL user, respectively.

\subsection{Ergodic WSR Analysis with Imperfect CSI} \label{Rate Analysis}
In the following, we derive the expression for the ergodic WSR $\mathcal{R}(\mathbb{E}_{\bmH|\hat{\bmH}})$ in the presence of estimated channel responses and CSI errors for the IRS-FD system with multi-antenna UL and DL users. Let $\widetilde{\bmU_k} = \bmU_k \bmU_k^H$ and $\widetilde{\bmV_j} = \bmV_j \bmV_j^H$ denote the transmit covariance matrices for UL user $k$ and DL user $j$, respectively. In the case of imperfect CSI, the received signal plus interference plus noise covariance matrices $\bmR_{k}$ and $\bmR_{j}$, including the CSI errors and the IRS phase response $\mathbf{\Theta}$, are given as
\begin{subequations} 
    \begin{equation}  \label{Ul_user_cov} 
    \begin{aligned}
        \bmR_{k} = & \overline{\bmH_k} \widetilde{\bmU_k} \overline{\bmH_k}^H + \overline{\bmH_0} \widetilde{\bmV_j} \overline{\bmH_0}^H + \sigma_0^2 \bmI,
    \end{aligned}
    \end{equation}
     \begin{equation} \label{Dl_user_cov} 
    \begin{aligned}
        \bmR_{j} = & \overline{\bmH_j} \widetilde{\bmV_j} \overline{\bmH_j}^H + \overline{\bmH}_{j,k} \widetilde{\bmU_k}   \overline{\bmH}_{j,k}^H   + \sigma_j^2 \bmI.
    \end{aligned}
    \end{equation}
\end{subequations}
The interference plus noise covariance matrices can be obtained as $\bmR_{\overline{k}} =  \bmR_{k} - \bmS_k, \bmR_{\overline{j}} =  \bmR_{j} - \bmS_j$, with $\bmS_k$ and $\bmS_j$ denoting the useful received signal covariance part.

Given the statistical distribution of the CSI errors \eqref{CE_modelling} and the channel estimates, the lower bound for the ergodic WSR $\mathcal{R}(\mathbb{E}_{\bmH|\hat{\bmH}})$ of an IRS-FD system with multi-antenna users with imperfect CSI can be approximated as
     
\begin{equation} \label{ach_rate_hat}
\begin{aligned}
\mathcal{R}(\mathbb{E}_{\bmH|\hat{\bmH}}) & =   w_k \mbox{ln}[\mbox{det}(\bmI +  \bmU_k^H (\hat{\bmH}_{k} +  \hat{\bmH}_{0,\theta} \mathbf{\Theta} \hat{\bmH}_{\theta,k})^H \mathbf{\Sigma}_{\overline{k}}^{-1} \\&  (\hat{\bmH}_{k} +  \hat{\bmH}_{0,\theta} \mathbf{\Theta} \hat{\bmH}_{\theta,k})  \bmU_k ) ]  +   w_j \mbox{ln}[\mbox{det}(\bmI +  \bmV_j^H (\hat{\bmH}_{j} + \\& \hat{\bmH}_{j,\theta}   \mathbf{\Theta} \hat{\bmH}_{\theta,0})^H \mathbf{\Sigma}_{\overline{j}}^{-1}    (\hat{\bmH}_{j} + \hat{\bmH}_{j,\theta}   \mathbf{\Theta} \hat{\bmH}_{\theta,0}) \bmV_j ) ],
\end{aligned}
\end{equation}

\begin{figure*}[!t]\small
\begin{subequations} \label{cov_true}
\begin{equation}  \label{cov_UL_user_k}
\begin{aligned}
    \mathbf{\Sigma}_{\overline{k}} =&   \mbox{Tr}(\widetilde{\bmU_k} \bmJ_k^T) \bmK_k      + \hat{\bmH}_{0,\theta} \mathbf{\Theta} \mbox{Tr}(\widetilde{\bmU_k} \bmJ_{\theta,k}^T) \bmK_{\theta,k} \mathbf{\Theta}^H \hat{\bmH}_{0,\theta}^H  + \mbox{Tr}(\mathbf{\Theta} \hat{\bmH}_{\theta,k} \widetilde{\bmU_k} \hat{\bmH}_{\theta,k} \mathbf{\Theta}^H \bmJ_{0,\theta}^T) \bmK_{0,\theta}   + \mbox{Tr}(\widetilde{\bmU_k} \bmJ_{\theta,k}^T) \mbox{Tr}(\mathbf{\Theta} \bmK_{\theta,k} \mathbf{\Theta}^H \bmJ_{0,\theta}^T) \bmK_{0,\theta} 
   \\& + \hat{\bmH}_0 \widetilde{\bmV_j} \hat{\bmH}_0^H  + \hat{\bmH}_0 \widetilde{\bmV_j} \bmH_{\theta,0}^H \mathbf{\Theta}^H \hat{\bmH}_{0,\theta} + \mbox{Tr}(\widetilde{\bmV_j} \bmJ_0^T) \bmK_0^T + \hat{\bmH}_{0,\theta} \mathbf{\Theta} \hat{\bmH}_{\theta,0} \widetilde{\bmV_j} \hat{\bmH}_0^H + \hat{\bmH}_{0,\theta} \mathbf{\Theta} \bmH_{\theta,0} \widetilde{\bmV_j} \bmH_{\theta,0}^H \mathbf{\Theta}^H \hat{\bmH}_{0,\theta}^H \\&+ \hat{\bmH}_{0,\theta} \mathbf{\Theta} \mbox{Tr}(\widetilde{\bmV_j} \bmJ_{\theta,0}^T) \bmK_{\theta,0} \mathbf{\Theta}^H \hat{\bmH}_{0,\theta}^H   + \mbox{Tr}(\mathbf{\Theta} \hat{\bmH}_{\theta,0} \widetilde{\bmV_j} \hat{\bmH}_{\theta,0}^H \mathbf{\Theta}^H \bmJ_{0,\theta}^T) \bmK_{0,\theta}  + \mbox{Tr}(\widetilde{\bmV_j} \bmJ_{\theta,0}^T) \mbox{Tr}(\mathbf{\Theta} \bmK_{\theta,0} \mathbf{\Theta}^H \bmJ_{0,\theta}^T) \bmK_{0,\theta} + \sigma_0^2 \bmI,
\end{aligned}   
\end{equation}
\begin{equation} 
    \begin{aligned}
       \mathbf{\Sigma}_{\overline{j}} =  &    \mbox{Tr}(\widetilde{\bmV_j} \bmJ_j^T) \bmK_j  + \hat{\bmH}_{j,\theta} \mathbf{\Theta} \mbox{Tr}(\widetilde{\bmV_j} \bmJ_{\theta,0}^T) \bmK_{\theta,0} \mathbf{\Theta}^H \hat{\bmH}_{j,\theta}^H  + \mbox{Tr}( \mathbf{\Theta} \hat{\bmH}_{\theta,0} \widetilde{\bmV_j} \hat{\bmH}_{\theta,0}^H \mathbf{\Theta}^H \bmJ_{j,\theta}^T ) \bmK_{j,\theta} + \mbox{Tr}(\widetilde{\bmV_j} \bmJ_{\theta,0}^T) \mbox{Tr}(\mathbf{\Theta} \bmK_{\theta,0} \mathbf{\Theta}^H \bmJ_{j,\theta}^T) \bmK_{j,\theta} \\& + \hat{\bmH}_{j,k} \widetilde{\bmU_k} \hat{\bmH}_{j,k}^H  +\hat{\bmH}_{j,k} \widetilde{\bmU_k} \hat{\bmH}_{\theta,k}^H \mathbf{\Theta}^H \hat{\bmH}_{j,\theta}^H + \mbox{Tr}(\widetilde{\bmU_k} \bmJ_{j,k}^T) \bmK_{j,k}   + \hat{\bmH}_{j,\theta} \mathbf{\Theta} \bmH_{\theta,k} \widetilde{\bmU_k} \hat{\bmH}_{j,k} + \hat{\bmH}_{j,\theta} \mathbf{\Theta} \bmH_{\theta,k} \widetilde{\bmU_k} \hat{\bmH}_{\theta,k}^H \mathbf{\Theta}^H \hat{\bmH}_{j,\theta}^H  \\& + \hat{\bmH}_{j,\theta} \mathbf{\Theta} \mbox{Tr}(\widetilde{\bmU_k} \bmJ_{\theta,k}^T) \bmK_{\theta,k} \mathbf{\Theta}^H \hat{\bmH}_{j,\theta}^H  + \mbox{Tr}(\mathbf{\Theta} \hat{\bmH}_{\theta,k} \widetilde{\bmU_k} \hat{\bmH}_{\theta,k}^H \mathbf{\Theta}^H \bmJ_{j,\theta}^T) \bmK_{j,\theta}  + \mbox{Tr}(\widetilde{\bmU_k} \bmJ_{\theta,k}^T ) \mbox{Tr}(\mathbf{\Theta} \bmK_{\theta,k}  \mathbf{\Theta}^H \bmJ_{j,\theta}^T )  \bmK_{j,\theta} + \sigma_j^2 \bmI. 
\end{aligned}
\end{equation}   \vspace{-4mm} \hrulefill  
\end{subequations}  
\end{figure*}
\noindent where $\mathbf{\Sigma}_{\overline{k}}$ and $\mathbf{\Sigma}_{\overline{j}}$ are given as in \eqref{cov_true}. This result is derived based on the identities: For any $\bmH\sim \mathcal{CN}(\hat{\bmH},\bmJ \otimes \bmK) $, there is $\mathbb{E}[\bmH \bmX \bmH^H ] = \hat{\bmH
} \bmX \hat{\bmH}^H + \mbox{Tr}(\bmX \bmJ^T) \bmK  $ and  $\mathbb{E}[\bmH^H \bmX \bmH] = \hat{\bmH
}^H \bmX \hat{\bmH} + \mbox{Tr}(\bmK\bmX) \bmJ^T$. Note that an upper bound for ergodic WSR can be easily derived from \eqref{cov_true} in the presence of ideal CSI, i.e., by setting $\bmJ$ and $\bmI$ equal to the identity.

\section{Robust Beamforming Via Expected WMMSE } \label{algorithmo}

The problem of ergodic WSR maximization \eqref{WSR_problem} is non-convex due to interference and in the case of perfect CSI it can be solved with an equivalent problem formulation of EWMMSE, which exploits the relationship between the WSR and the expected error covariance matrices \cite{negro2012sum}. 


\subsection{Digital Combining}
We assume that the FD BS for UL user $k$ and the DL user $j$  apply the combiners $\bmF_k$ and $\bmF_j$ to estimate their data streams as $\hat{\bms}_k = \bmF_k \bmy_k$ and $\hat{\bms}_j = \bmF_j \bmy_j$.
Let $\bmE_{\tilde{k}}$ and $\bmE_{\tilde{j}}$ denote the MSE error matrices for instantaneous CSI for UL user $k$ and DL user $j$, respectively, which can be written as $\bmE_{\tilde{k}} =  \mathbb{E}[\bme_k \bme_k^H], \quad  \bmE_{\tilde{j}} =  \mathbb{E}[\bme_j \bme_j^H]$
where $\bme_k = \bmF_k \bmy_k - \hat{\bms}_k$ and $\bme_j =\bmF_j \bmy_j - \hat{\bms}_j$ denote the error vectors. Let $\bmQ_k, \bmT_j, \bmT_{0}$ and $\bmQ_{j,k}$ denote the matrices defined as
\begin{subequations} \label{aux_Q}
    \begin{equation}
        \bmQ_k  =  \mathbb{E}_{\bmH|\hat{\bmH}}(\overline{\bmH}_k \widetilde{\bmU_k} \overline{\bmH}_k^H), \,  
         \bmT_j =  \mathbb{E}_{\bmH|\hat{\bmH}}(\overline{\bmH}_j \widetilde{\bmV_j} \overline{\bmH}_j^H),
    \end{equation}
    \begin{equation}
       \bmT_{0} = \mathbb{E}_{\bmH|\hat{\bmH}}(\overline{\bmH}_0 \widetilde{\bmV_j} \overline{\bmH}_0^H),  \,
        \bmQ_{j,k} =   \mathbb{E}_{\bmH|\hat{\bmH}}(\overline{\bmH}_{j,k}  \widetilde{\bmU_k} \overline{\bmH}_{j,k}^H).
    \end{equation}
\end{subequations}
Since we adopt the EWMMSE approach, consider the expected MSE of the UL and DL user as
\begin{subequations} \label{error_cov_comb}
    \begin{equation}
    \begin{aligned}
        \bmE_k & =   \mathbb{E}_{\bmH|\hat{\bmH}}[\bmE_{\tilde{k}}] \\ &= \bmF_k \bmQ_k \bmF_k^H - \bmF_k 
        (\hat{\bmH}_k + \hat{\bmH}_{0,\theta} \mathbf{\Theta} \hat{\bmH}_{\theta,k})  \bmU_k  
         + \bmF_k \bmT_{0} \bmF_k^H  \\& \quad + \sigma_0^2 \bmF_k \bmF_k^H - \bmU_k^H \hat{\bmH}_k^H \bmF_k^H   - \bmU_k^H \hat{\bmH}_{\theta,k}^H  \mathbf{\Theta}^H \hat{\bmH}_{0,\theta}^H \bmF_k^H - \bmI.
    \end{aligned}
    \end{equation}
        \begin{equation}
    \begin{aligned}
        \bmE_j  & =   \mathbb{E}_{\bmH|\hat{\bmH}}[\bmE_{\tilde{j}}] \\ &=
        \bmF_j  \bmT_j \bmF_j^H - \bmF_j (\hat{\bmH}_j + 
        \hat{\bmH}_{j,\theta}  \mathbf{\Theta} \hat{\bmH}_{\theta,0} )  \bmV_j 
      + \bmF_j  \bmQ_{j,k} \bmF_j^H \\& \quad  + \sigma_j^2 \bmF_j \bmF_j^H  - \bmV_j^H  \hat{\bmH}_j^H \bmF_j^H  -
       \bmV_j^H\hat{\bmH}_{\theta,0}^H \mathbf{\Theta}^H  \hat{\bmH}_{j,\theta}^H   \bmF_j^H - \bmI.
    \end{aligned}
    \end{equation}
\end{subequations}

By minimizing the trace of the \eqref{error_cov_comb}, the optimal EWMMSE combiners can be derived as
  
\begin{subequations} \label{MMSE_comb}
\begin{equation}
    \bmF_k = \bmU_k^H (\hat{\bmH}_k^H + \hat{\bmH}_{\theta,k}^H \mathbf{\Theta}^H \hat{\bmH}_{0,\theta}^H ) ( \bmQ_k +  \bmT_{0} + \sigma_0^2 \bmI )^{-1},
\end{equation}
\begin{equation}
    \bmF_j = \bmV_j^H (\hat{\bmH}_j^H + \hat{\bmH}_{\theta,0}^H \mathbf{\Theta}^H \hat{\bmH}_{j,\theta}^H ) (  \bmT_j  +\bmQ_{j,k} + \sigma_j^2 \bmI )^{-1}.
\end{equation}
\end{subequations}




\subsection{Active Digital Beamforming Under Imperfect CSI}

Given the optimal combiners, the EWMMSE problem with respect to the digital beamformers under the average total sum-power constraint and given the IRS phase response $\mathbf{\Theta}$ fixed, can be formally stated as 

\begin{subequations}\label{dig_BF_problem}
\begin{equation}
     \underset{\substack{\bmV_j,\bmU_k}}{\min} \quad  \mbox{Tr}(\bmW_k \bmE_k) +   \mbox{Tr}( \bmW_j \bmE_j), \vspace{-2mm}
\end{equation} \label{WSR}
\begin{equation} \vspace{-2mm} \label{pow_constraints}
  \text{s.t.}  \;  \mbox{Tr}( \bmU_k \bmU_k^H ) 	\preceq \alpha_k,\; \&
  \;\;  \mbox{Tr} (  \bmV_j  \bmV_j^H ) \leq  \alpha_0, 
\end{equation}
\end{subequations}
where $\bmW_i$ is a constant weight matrix associated with node $i$. The problem \eqref{dig_BF_problem} and the WSR maximization problem \eqref{WSR_problem} are equivalent if their gradient results to be the same, which can be assured if the weight matrices are chosen as 
\begin{equation} \label{W_matrix}
        \bmW_k = \frac{w_k}{\mbox{ln}\; 2} ( \bmE_k )^{-1}, \quad
        \bmW_j = \frac{w_j}{\mbox{ln}\;2} ( \bmE_j )^{-1}.
\end{equation}
Such a result can be easily shown by following a similar proof provided in \cite[Appendix A]{cirik2015weighted}, but carried out for the case of average MSE error under the imperfect CSI case. To optimize the digital beamformers $\bmV_j,\bmU_k$
we take the partial derivative of the Lagrangian function of 
\eqref{dig_BF_problem} with respect to their conjugate, which leads to the following optimal beamformers   

\begin{figure*}\small
\begin{subequations}
    \begin{equation} \label{X_k}
    \begin{aligned}
        \bmX_{k} = & (\hat{\bmH}_k^H + \hat{\bmH}_{\theta,k}^H \mathbf{\Theta}^H \hat{\bmH}_{0,\theta}^H) \bmF_k^H \bmW_k \bmF_k \hat{\bmH}_k + \mbox{Tr}(\bmK_k \bmF_k^H \bmW_k \bmF_k)  \bmJ_k^T + \hat{\bmH}_{\theta,k}^H \mathbf{\Theta}^H \hat{\bmH}_{0,\theta}^H \bmF_k^H \bmW_k \bmF_k \hat{\bmH}_{0,\theta} \mathbf{\Theta} \hat{\bmH}_{\theta,k} \\& + \bmH_k^H \bmF_k^H \bmW_k \bmF_k \hat{\bmH}_{0,\theta} \mathbf{\Theta} \hat{\bmH}_{\theta,k}  +
        \mbox{Tr}(\bmK_{\theta,k} \mathbf{\Theta}^H \hat{\bmH}_{0,\theta}^H \bmF_k^H \bmW_k \bmF_k \hat{\bmH}_{0,\theta} \mathbf{\Theta} )  \bmJ_{\theta,k}^T
         + \hat{\bmH}_{\theta,k}^H \mathbf{\Theta}^H
         \mbox{Tr}( \bmK_{0,\theta}\bmF_k^H \bmW_k \bmF_k )  \bmJ_{0,\theta}^T \mathbf{\Theta} \hat{\bmH}_{\theta,k}   \\&
        + \hat{\bmH}_{j,k}^H \bmF_j^H \bmW_j \bmF_j \hat{\bmH}_{j,\theta} \mathbf{\Theta} \hat{\bmH}_{\theta,k} 
            + (\hat{\bmH}_{j,k}^H + \hat{\bmH}_{\theta,k}^H \mathbf{\Theta}^H \hat{\bmH}_{j,\theta}^H )\bmF_j^H \bmW_j \bmF_j \hat{\bmH}_{j,k}  
           + \mbox{Tr}(\bmK_{j,k} \bmF_j^H \bmW_j \bmF_j )  \bmJ_{j,k}^T
           \\&+ \mbox{Tr}(\bmK_{0,\theta} \bmF_k^H \bmW_k \bmF_k) \mbox{Tr}(\bmK_{\theta,k} \mathbf{\Theta}^H \bmJ_{0,\theta}^T \mathbf{\Theta}) \bmJ_{\theta,k}^T + \hat{\bmH}_{\theta,k}^H \mathbf{\Theta}^H \hat{\bmH}_{j,\theta}^H \bmF_j^H \bmW_j \bmF_j \hat{\bmH}_{j,\theta} \mathbf{\Theta} \hat{\bmH}_{\theta,k}  + 
           \mbox{Tr}(\bmK_{\theta,k} \mathbf{\Theta}^H \hat{\bmH}_{j,\theta}^H \bmF_j^H \bmW_j \bmF_j \hat{\bmH}_{j,\theta} \mathbf{\Theta} ) \bmJ_{\theta,k}^T  
           \\&  + \hat{\bmH}_{\theta,k}^H \mathbf{\Theta}^H 
            \mbox{Tr}(\bmK_{j,\theta} \bmF_j^H \bmW_j \bmF_j )
            \bmJ_{j,\theta}^T  \mathbf{\Theta} \hat{\bmH}_{\theta,k}
            +   \mbox{Tr}(\bmK_{j,\theta}  \bmF_j^H \bmW_j \bmF_j) \mbox{Tr}(\bmK_{\theta,k} \mathbf{\Theta}^H \bmJ_{j,\theta}^T \mathbf{\Theta}) \bmJ_{\theta,k}^T,
     \end{aligned}
    \end{equation}
    \begin{equation} \label{X_j}
    \begin{aligned}
                \bmX_j = & (\hat{\bmH}_{j}^H + \hat{\bmH}_{\theta,0}^H \mathbf{\Theta}^H \hat{\bmH}_{j,\theta}^H) \bmF_j^H \bmW_j \bmF_j \hat{\bmH}_j + \mbox{Tr}(\bmK_j \bmF_j^H \bmW_j \bmF_j ) \bmJ_{j}^T
                  + \hat{\bmH}_{\theta,0}^H \mathbf{\Theta}^H \hat{\bmH}_{j,\theta}^H \bmF_j^H \bmW_j \bmF_j \hat{\bmH}_{j,\theta} \mathbf{\Theta} \hat{\bmH}_{\theta,0}  + \hat{\bmH}_j^H \bmF_j^H \bmW_j \bmF_j \hat{\bmH}_{j,\theta} \mathbf{\Theta} \hat{\bmH}_{\theta,0}\\& 
                  + \mbox{Tr}(\bmK_{\theta,0}  \mathbf{\Theta}^H \hat{\bmH}_{j,\theta}^H  \bmF_j^H \bmW_j \bmF_j \hat{\bmH}_{j,\theta} \mathbf{\Theta} ) \bmJ_{\theta,0}^T
                  + \hat{\bmH}_{\theta,0}^H \mathbf{\Theta}^H 
                   \mbox{Tr}(\bmK_{j,\theta} \bmF_j^H \bmW_j \bmF_j ) \bmJ_{j,\theta}^T \mathbf{\Theta} \hat{\bmH}_{\theta,0}
                + \bmH_0^H \bmF_k^H \bmW_k \bmF_k \hat{\bmH}_{0,\theta} \mathbf{\Theta} \hat{\bmH}_{\theta,0}
                \\&
                + (\hat{\bmH}_0^H  + \hat{\bmH}_{\theta,0}^H \mathbf{\Theta}^H \hat{\bmH}_{0,\theta}^H ) \bmF_k^H \bmW_k \bmF_k \hat{\bmH}_0 + \mbox{Tr}(\bmK_{0} \bmF_k^H \bmW_k \bmF_k)  \bmJ_{0}^T  + \mbox{Tr}(\bmK_{j,\theta} \bmF_j^H \bmW_j \bmF_j )  \mbox{Tr}(\bmK_{\theta,0} \mathbf{\Theta}^H \bmJ_{j,\theta}^T  \mathbf{\Theta} ) \bmJ_{\theta,0}^T   \\&  
                + \hat{\bmH}_{\theta,0}^H \mathbf{\Theta}^H \hat{\bmH}_{0,\theta}^H \bmF_k^H \bmW_k \bmF_k \hat{\bmH}_{0,\theta} \mathbf{\Theta} \hat{\bmH}_{\theta,0} + \mbox{Tr}(\bmK_{\theta,0} \mathbf{\Theta}^H \hat{\bmH}_{0,\theta}^H \bmF_k^H \bmW_k \bmF_k \hat{\bmH}_{0,\theta} \mathbf{\Theta} ) \bmJ_{\theta,0}^T
                + \hat{\bmH}_{\theta,0}^H \mathbf{\Theta}^H \mbox{Tr}(\bmK_{0,\theta} \bmF_k^H \bmW_k \bmF_k )  \bmJ_{0,\theta}^T \mathbf{\Theta} \hat{\bmH}_{\theta,0} \\& + 
                \mbox{Tr}(\bmK_{0,\theta} \bmF_k^H \bmW_k \bmF_k) \mbox{Tr}(\bmK_{\theta,0} \mathbf{\Theta}^H \bmJ_{0,\theta}^T \mathbf{\Theta}) \bmJ_{\theta,0}^T.
                \end{aligned}
    \end{equation} \vspace{-5mm} \hrulefill
    \end{subequations}
\end{figure*}

\begin{subequations}
\begin{equation} \label{UL_BF}
    \bmU_k = (\bmX_k + \lambda_k \bmI)^{-1} (\hat{\bmH}_{k}^H + \hat{\bmH}_{\theta,k}^H \mathbf{\Theta}^H \hat{\bmH}_{0,\theta}^H) \bmF_k^H \bmW_k,
\end{equation}
\begin{equation} \label{DL_BF}
    \bmV_j = (\bmX_j + \lambda_0 \bmI)^{-1} (\hat{\bmH}_{j}^H + \hat{\bmH}_{\theta,0}^H \mathbf{\Theta}^H \hat{\bmH}_{j,\theta}^H) \bmF_j^H \bmW_j,
\end{equation}
\end{subequations}
where $\bmX_k$ and $\bmX_j$ are defined in \eqref{X_k} and \eqref{X_j}, respectively, and the scalars $\lambda_k$ and $\lambda_0$ denote the Lagrange multiplier for the uplink user $k$ and the FD BS. The multipliers can be searched while performing power allocation for the users given the average total sum-power constraints. Namely, consider the singular value decomposition (SVD) of the matrices as $\bmX_{k} = \bmA_k \mathbf{\Lambda}_k \bmB_k $  and $\bmX_{j}=\bmA_j \mathbf{\Lambda}_j \bmB_j$, where $\bmA_i$ and $\bmB_i$ denote the left and right unitary matrices obtained with SVD and $\mathbf{\Lambda}_i$ denote the singular values. The average power constraints \eqref{pow_constraints}, after some simplifications, can be written as 
\begin{subequations}
\begin{equation}
    \mbox{Tr}(\bmV_j \bmV_j^H) = \frac{\sum_{i=i}^{M_0} \bmS_j(i,i)} {(\lambda_0 + \mathbf{\Lambda}_j(i,i))^2}, 
\end{equation}
\begin{equation}
    \mbox{Tr}(\bmU_k \bmU_k^H) = \frac{\sum_{i=i}^{M_k} \bmS_k(i,i)} {(\lambda_j + \mathbf{\Lambda}_k(i,i))^2},
\end{equation}
\end{subequations}
where the matrices $\bmS_k$ and $\bmS_j$ are defined as 
\begin{subequations}
\begin{equation}
\begin{aligned}
     \bmS_j= \bmB_j (\hat{\bmH}_{j} + \hat{\bmH}_{j,\theta}  \mathbf{\Theta} \hat{\bmH}_{\theta,0})^H & \bmF_j^H \bmW_j \bmW_j \bmF_j \\& (\hat{\bmH}_{j} + \hat{\bmH}_{j,\theta}   \mathbf{\Theta} \hat{\bmH}_{\theta,0})\bmA_j,
\end{aligned}
\end{equation}
\begin{equation}
\begin{aligned}
     \bmS_k = \bmB_k (\hat{\bmH}_{k} +  \hat{\bmH}_{0,\theta}  \mathbf{\Theta} \hat{\bmH}_{\theta,k})^H & \bmF_k^H \bmW_k \bmW_k \bmF_k \\& (\hat{\bmH}_{k} +  \hat{\bmH}_{0,\theta} \mathbf{\Theta} \hat{\bmH}_{\theta,k}) \bmA_k.
\end{aligned}
\end{equation}
\end{subequations}
The optimal Lagrange multipliers satisfying the average total power constraints can be searched with a linear search, and in this work, we adopt the Bisection method. If the obtained values of the multipliers are negative, then we replace them with zero. Note that the average power constraint is met based on the CSI estimates.

\subsection{Passive Beamforming Under Imperfect CSI}
In this section, we consider optimizing the phase response of the IRS to jointly assist the UL and DL channels for rate enhancement. Let $\bmS, \bmT$ and $\bmZ$ denote the matrices independent of the IRS phase response, given in Appendix \ref{IRS-Opt-matrices}.
The EWMMSE optimization problem for the IRS phase response $\mathbf{\Theta}$ under imperfect CSI, given the matrices $\bmS, \bmT$ and $\bmZ$, can be formally stated as

 \begin{subequations} \label{full_problem_ir} 
\begin{equation} \label{IRS_r_opt_problem}
    \underset{\substack{\mathbf{\Theta} }}{\min} \quad \mbox{Tr}(\mathbf{\Theta}^H \bmZ \mathbf{\Theta} \bmT ) +\mbox{Tr}(\mathbf{\Theta}^H \bmS^H ) + \mbox{Tr}(\mathbf{\Theta} \bmS ) + c,
    \end{equation}
\begin{equation} \label{c1_r}
\text{s.t.}  |\bm{\theta}(i)|=1,  \quad \forall i,
\end{equation}
\end{subequations}
where the scalar $c$ denotes the constant terms, independent of $\mathbf{\Theta}$. In \eqref{IRS_r_opt_problem}, the problem is stated with respect to the matrix $\mathbf{\Theta}$. However, we wish to maximize only the diagonal response of such matrix to maximize the ergodic WSR or minimize the expected MSE, because the off-diagonal elements result to be zero. Therefore, we first consider restating the problem \eqref{IRS_r_opt_problem} with respect to $\bm{\theta}$, being a vector made of the diagonal elements of $\mathbf{\Theta}$. For doing so, we write the trace terms as
\begin{subequations}
    \begin{equation}
    \mbox{Tr}(\mathbf{\Theta}^H \bmZ \mathbf{\Theta} \bmT ) = \bm{\theta}^H  \mathbf{\Sigma} \bm{\theta}, \quad \mbox{where}\; \mathbf{\Sigma} = \bmZ \odot \bmT^T,
\end{equation}
\begin{equation}
    \mbox{Tr}(\mathbf{\Theta}^H \bmS^H ) = {\bms}^H \bm{\theta}^*,\quad \mbox{Tr}(\mathbf{\Theta} \bmS ) = {\bms}^T \bm{\theta},
\end{equation}
\end{subequations}
where $\mathbf{s}$ denotes the vector containing only the diagonal elements of the matrix $\bmS$. Problem \eqref{IRS_r_opt_problem} can be restated with respect to the vector $\bm{\theta}$ as 
\begin{subequations} \label{refprob}
\begin{equation} \label{IRS_r_opt_restated_2}
    \underset{\substack{\bm{\theta}}}{\min} \quad \bm{\theta}^H  \mathbf{\Sigma} \bm{\theta} +{\bms_r}^H \bm{\theta}^* + {\bms}^T \bm{\theta}_l , \vspace{-2mm}
    \end{equation}
\begin{equation} \label{c1_rest}
\text{s.t.} \quad 
   |\bm{\theta}(i)| = 1,  \quad \forall i. \vspace{-2mm}
\end{equation}
\end{subequations}
Problem \eqref{refprob} is still very challenging as it is a non-convex problem due to the unit-modulus constraint. To solve it we adopt the majorization-maximization method \cite{pan2020multicell}. Such method aims to solve a more difficult problem by constructing a series of more tractable problems stated with the upper bound. Let $\mathcal{R}(\bm{\theta}^{(n)})$ denote the function evaluating \eqref{refprob} at the $n$-th iteration for computed $\bm{\theta}$. Let $\mathcal{R}_u(\bm{\theta}|\bm{\theta}^{(n)})$ denote an upper bound constructed at $n$-th iteration for $\mathcal{R}$. According to \cite{pan2020multicell}, for the problem of the form \eqref{refprob}, an upper bound can be constructed as

\begin{equation} \label{UB}
     \mathcal{R}_u(\bm{\theta}|\bm{\theta}^{(n)}) = 2 \mbox{Re}\{{\bms}^H \bmq^{(n)}\} + c_u, \vspace{-1mm}
 \end{equation}
where $c_u$  denote constant terms in the upper bound independent of $\bm{\theta}$ and $\bmq^{(n)}$ is given by 

\begin{equation} \label{qn_calcolo}
    \bmq^{(n)} = (\lambda^{max} \bmI - \mathbf{  \Sigma}) \bm{\theta}^{(n)} - {\bms}^*,
\end{equation}
with $\lambda^{max}$ denoting the maximum eigenvalues of $\mathbf{\Sigma}$. Based on the result above, the hard non-convex optimization problem \eqref{IRS_r_opt_restated_2} simplifies to a series of the following problem  
 \begin{subequations}  
\begin{equation} \label{IRS_r_optimized_ert}
    \underset{\substack{\bm{\theta}}}{\min} \quad  2 \mbox{Re}\{{\bms}^H \bmq^{(n)}\},   \vspace{-3mm}
    \end{equation}
\begin{equation} \label{c1_trew}  \vspace{-3mm}
\text{s.t.} \quad 
   |\bm{\theta}(i)| = 1,  \quad \forall i,
   \end{equation}
\end{subequations}
which need to be solved iteratively until convergence for each update of $\bm{\theta}$. By solving \eqref{IRS_r_optimized_ert}, we get the optimal solution of $\bm{\theta}$ at $n+1$-th iteration, given the $\bm{\theta}$ at the $n$-th iteration as
 
\begin{equation} \label{solution_phi_r}
    \bm{\theta}^{(n+1)} = e^{i \angle\bmq^{(n)}}. 
\end{equation}
Formally, the optimization algorithm to optimize the IRS phase response is given in Algorithm \ref{IRS_opt} and the joint optimization procedure is provided in Algorithm \ref{alg_1}.

\begin{algorithm}[t]  \scriptsize
\caption{Optimization of IRS Phase Response }\label{alg_1}
\textbf{Initialize:}   $\bm{\theta}^0$, iteration index $n=1$, accuracy $\epsilon$.\\
\textbf{Evaluate:} $\mathcal{R}(\bm{\theta}^0)$.\\
\textbf{Repeat until convergence}
\begin{algorithmic}
\STATE \hspace{0.001cm} Calculate $\bmq_i^{(n)}$ with \eqref{qn_calcolo}.
\STATE  \hspace{0.001cm} Update $\bm{\theta}_i^{(n+1)}$ with  \eqref{solution_phi_r}.\\
\STATE  \hspace{0.001cm} \textbf{if} $|\mathcal{R}^{(n+1)} - \mathcal{R}(\bm{\theta}_i^{(n)})|/\mathcal{R}(\bm{\theta}_i^{(n+1)}) \leq \epsilon$
\STATE  \hspace{0.4cm} Stop and return $\bm{\phi}_i^{(n+1)}$.
\end{algorithmic}   \label{IRS_opt} \vspace{-1mm}
\end{algorithm}

\begin{algorithm}[t]   \scriptsize
\caption{Robust Beamforming for MIMO IRS-FD system}
\textbf{Initialize} the iteration index $n$, accuracy $\epsilon$, beamformers and combiners.\\
\textbf{Repeat until convergence}
\begin{algorithmic}
\STATE for $i$, where $i =k $ or $i=j$\\
\STATE \hspace{0.3cm} Update $\bmF_i$ with \eqref{MMSE_comb}.\\
\STATE \hspace{0.3cm} Update $\bmW_i$ with \eqref{W_matrix}.\\
\STATE \hspace{0.3cm} \textbf{if} i=k $\rightarrow$ update $\bmU_k$ with \eqref{UL_BF}.\\
\STATE \hspace{0.3cm} \textbf{else} $\rightarrow$  update $\bmV_j$ with \eqref{DL_BF}.\\
\STATE  Update $\mathbf{\Theta}$ with Algorithm  \ref{alg_1}.\\
\STATE  \textbf{if} convergence condition is satisfied\\
\STATE  \hspace{0.4cm} Stop and return the optimized variables. 
\end{algorithmic} \label{alg_1} \vspace{-1.5mm}
\end{algorithm}  
Note that each update of the beamformers leads to a monotonic decrease in the EWMMSE which assures the convergence of the proposed methods. A more formal proof can be presented by following a similar strategy as \cite{christensen2008weighted}. However, due to space limitations, we omit it here, which will be available in the extended version of the paper.

\section{Numerical Results} \label{risultati}
In this section, we present extensive simulation results to evaluate the performance of the proposed robust beamforming design. We consider the FD BS and the IRS to be centered in the positions $(0\mbox{m},0\mbox{m},0\mbox{m})$ and $(20\mbox{m},10\mbox{m},0\mbox{m})$, respectively, in the three-dimensional coordinates. The UL and the DL user are assumed to be randomly located in circles of radius $r=8$~m centered in the positions $(20\mbox{m},0\mbox{m},30\mbox{m})$ and $(30\mbox{m},0\mbox{m},20\mbox{m})$, respectively. We assume that the FD BS and the users are equipped with the uniform linear array (ULA), placed at the distance of half-wavelength. The IRS of $10 \times 10 = 100$ elements is assumed to assist the MIMO FD communication unless otherwise stated.  The MIMO FD BS is assumed to have $M_0=15$ transmit and $N_0=8$ receive antennas. The UL user $k$ and the DL user $j$ is assumed to have $M_k=5$ transmit and $5$ receive antennas, respectively, and served with $d_k=d_j=2$ data streams.
The large scale path loss is modelled as \cite{pan2020multicell} and the SI channel is modelled as a Rician fading \cite{sheemar2021hybrid_9} with Rician factor $1$. The direct links between the users and the FD BS are also modelled with a Rician fading channel model with a Rician factor of $1$. The channels involving the IRS are modelled according to Rayleigh fading \cite{guo2019weighted}.
We define the transmit signal-to-noise (SNR) of our system as $SNR = \frac{\alpha_0}{\sigma_j^2} = \frac{\alpha_k}{\sigma_0^2}$,
where $\alpha_0$ and $\alpha_k$ are the average total transmit power at the FD BS and the multi-antenna UL user $k$. We assume that the CSI errors to be i.i.d zero-mean  circularly
symmetric complex Gaussian distribution with the same variance $\sigma_{csi}^2$, and therefore, the error covariance matrices set chosen as $\bmJ_i = \bmI$ and $\bmK_{j} = \sigma_{csi}^2 \bmI$, $\forall i,j$. Since the variance of the CSI errors strictly depend on the average transmit power (used also to estimate the channels)  and the noise variance, we assume that $\sigma_{csi}^2$ decays as $\mathcal{O}(SNR^{-\alpha})$, for some constant $\alpha$, satisfying the CSI error decay rate inversely proportional to the SNR. As the SNR represents the transmit SNR, note that $SNR \rightarrow \infty$ is equivalent to $\alpha_k=\alpha_0 \rightarrow \infty$, which enhances the CSI quality and reduces the CSI error variance as $\sigma_{csi}^2 \rightarrow 0$.  The CSI errors variance is set as $\sigma_{csi}^2 = \rho /SNR^\alpha$, where $\rho$ denotes a scale factor and $\alpha \in [0,1]$ determines the quality of the CSI. Maintaining the quality of the CSI with $\alpha=1$ could be exhausting in terms of resources required to acquire the CSI and therefore we set $\alpha = 0.6$.
 
\begin{figure}[t]
     \centering
\includegraphics[width=0.7\columnwidth,height=4cm]{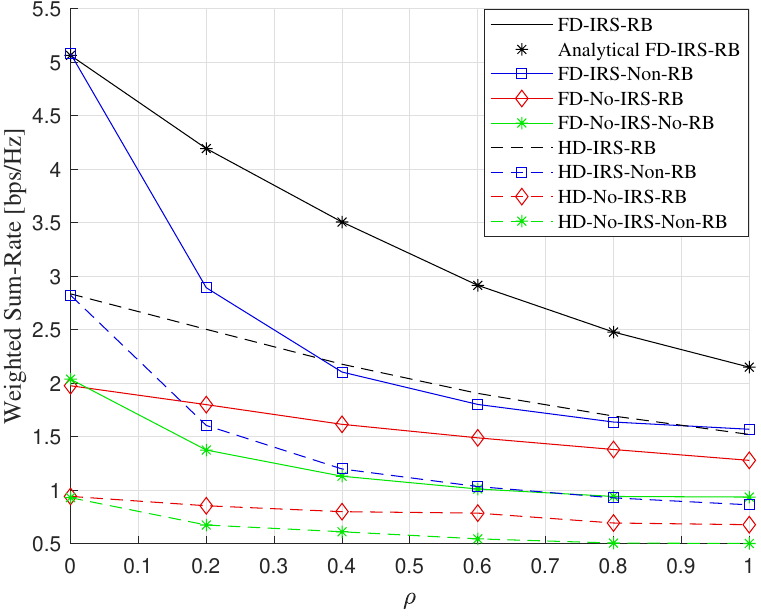}\vspace{-2.5mm}
    \caption{Average WSR as a function of $\rho$  with SNR$=30~$dB.} \label{WSR_roh} \vspace{-6mm}
\end{figure}

We label our proposed design as a \emph{FD-IRS-RB}.
For comparison, we define the following benchmark schemes: 1) \emph{FD-IRS-Non-RB}: FD system assisted with IRS and non-robust beamforming, i.e., CSI is treated as perfect; 2)  \emph{FD-No-IRS-RB}: FD system with no IRS and robust beamforming; 3) \emph{FD-No-IRS-Non-RB}: FD system with no IRS and non-robust beamforming; 4)\emph{HD-IRS-RB}: HD system assisted with IRS and robust beamforming; 5) \emph{HD-IRS-Non-RB}: HD system assisted with IRS and non-robust beamforming; 6) \emph{HD-No-IRS-RB}: HD system with no IRS and robust beamforming; 7) \emph{HD-No-IRS-Non-RB}: HD system with no IRS and non-robust beamforming.
We also compare our approximation proposed for the analytical ergodic WSR derived above, which is labelled as \emph{Analytical FD-IRS-RB} and compare it with the ergodic rate achieved with the EWMMSE robust beamforming approach. 
\begin{figure}
   \centering
\includegraphics[width=0.7\columnwidth,height=4cm]{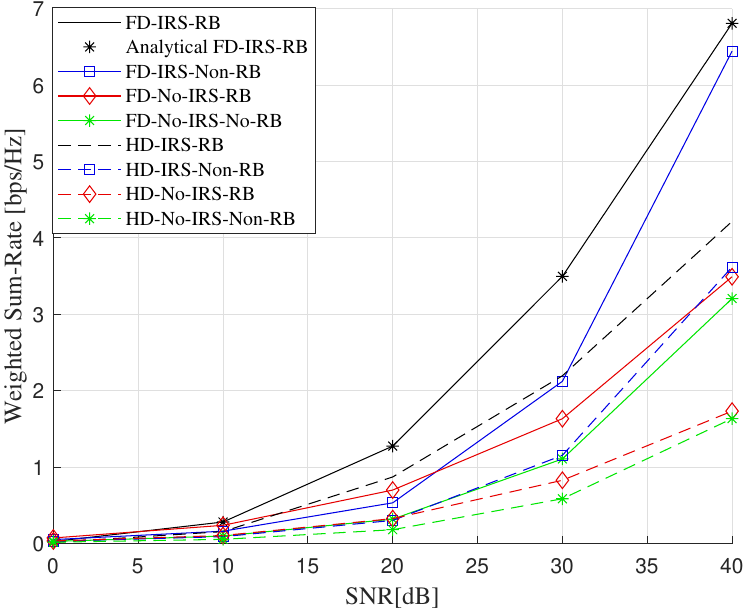}  \vspace{-1.5mm}
    \caption{Average WSR as a function of SNR with $\rho=0.4$.}   \label{WSR_roh_04} \vspace{-8mm}
\end{figure}


Fig. \ref{WSR_roh} shows the performance of the proposed robust joint beamforming design as a function of the scale factor $\rho$ dictating the CSI error variance, by means of Monte Carlo simulations at SNR$=30~$dB as a function of $\rho$. It is shown that the proposed robust design achieves significant performance gain compared to the naive FD-IRS-Non-RB scheme, which does not account for the CSI errors. Moreover, we can also see that the achieved ergodic WSR accurately matches the analytical result derived above. It is to be noted that as the CSI error variance gets extremely large, the proposed statistically robust beamforming design preserves significant robustness against the uncertainties in the CSI quality.

Fig. \ref{WSR_roh_04} shows the ergodic WSR of the proposed design as a function of the transmit SNR with $\rho =0.4$. We can see that our robust design achieves significant gain in the presence of CSI errors compared to the other schemes. Theoretically, the FD systems offer a twofold gain in the WSR compared to the HD systems. However, Fig.~\ref{WSR_roh_04} for the scheme FD-IRS-Non-RB, at low SNR, i.e., high CSI error variance, shows that the IRS-FD system may achieve no gain compared to the IRS-aided HD system  at any SNR level if non-robust beamforming approach is adopted. This is due to the fact that the residual SI can become extremely large in the presence of large CSI errors and can significantly degrade the performance. However, we can observe that as the CSI error variance decreases, the performance of the non-robust beamforming scheme tends toward the performance of the robust beamforming schemes, which tends towards the two-fold improvement in the WSR due to quasi-ideal simultaneous transmission and reception.

\section{Conclusions} \label{Conc} 
In this paper, a novel and statistically robust beamforming design for an IRS-FD system under imperfect CSI is presented. A closed-form approximation for the ergodic WSR is first derived given the statistical distribution of the errors. Then the ergodic WSR maximization problem is handled based on the EWMMSE method leading to two layers of sub-problems which are solved iteratively based on alternating optimization. Simulation results show that imperfect CSI can potentially limit the performance of the IRS-FD systems, and adopting robust beamforming leads to significant improvement in the performance gains.

\vspace{-2mm}
\appendix
\vspace{-2mm}
 \section{Matrices to optimize \mathbf{\Theta}} \small \label{IRS-Opt-matrices}
\begin{equation} 
 \begin{aligned}
     \bmS = & \hat{\bmH}_{0,\theta}^H \bmF_k^H (\bmW_k \bmF_k \hat{\bmH}_k \widetilde{\bmU_k} \hat{\bmH}_{\theta,k}^H +  \bmW_k \bmF_k \hat{\bmH}_0 \widetilde{\bmV_j} \hat{\bmH}_{\theta,0}^H  \\& + \bmW_k \bmU_k^H \hat{\bmH}_{\theta,k}^H ) +  \hat{\bmH}_{j,\theta}^H \bmF_j^H (\bmW_j \bmF_j \bmH_j \widetilde{\bmV_j} \bmH_{\theta,0}^H  \\& + \bmW_j \bmF_j \bmH_{j,k} \widetilde{\bmU_k} \bmH_{\theta,k}^H - \bmW_j \bmV_j^H \hat{\bmH}_{\theta,0}^H ),
 \end{aligned}
 \end{equation}

 \begin{equation}
 \begin{aligned} \footnotesize
       \bmZ = & \hat{\bmH}_{0,\theta}^H \bmF_k^H \bmW_k \bmF_k \bmH_{0,\theta} + \hat{\bmH}_{0,\theta}^H \bmF_k^H \bmW_k \bmF_k  \bmH_{0,\theta} \\& + 
       \mbox{Tr}(\bmF_k^H \bmW_k \bmF_k \bmJ_{0,\theta}^T) \bmK_{0,\theta} +  \mbox{Tr}( \bmF_k^H \bmW_k \bmF_k \bmJ_{0,\theta}^T) \bmK_{0,\theta} \\& + \mbox{Tr}(\bmF_k^H \bmW_k \bmF_k \bmJ_{0,\theta}^T) \bmK_{0,\theta} + \hat{\bmH}_{0,\theta}^H \bmF_k^H \bmW_k \bmF_k \hat{\bmH}_{0,\theta} \\& + \hat{\bmH}_{0,\theta}^H \bmF_k^H \bmW_k \bmF_k \hat{\bmH}_{0,\theta} + 
       \mbox{Tr}( \bmF_k^H \bmW_k \bmF_k \bmJ_{0,\theta}^T) \bmK_{0,\theta} \\& +\hat{\bmH}_{j,\theta}^H \bmF_j \bmW_j \bmF_j \hat{\bmH}_{j,\theta} + \hat{\bmH}_{j,\theta}^H \bmF_j^H \bmW_j \bmF_j \hat{\bmH}_{j,\theta} \\& + 
       \mbox{Tr}(\bmF_j^H \bmW_j \bmF_j \bmJ_{j,\theta}^T) \bmK_{j,\theta}
        + \mbox{Tr}(\bmF_j^H \bmW_j \bmF_j \bmJ_{j,\theta}^T) \bmK_{j,\theta} \\& +
         \hat{\bmH}_{j,\theta}^H \bmF_j^H \bmW_j \bmF_j \hat{\bmH}_{j,\theta} +  \hat{\bmH}_{j,\theta}^H \bmF_j^H \bmW_j \bmF_j \hat{\bmH}_{j,\theta} \\& +
         \mbox{Tr}(\bmF_j^H \bmW_j \bmF_j \bmJ_{j,\theta}^T) \bmK_{j,\theta} +  
         \mbox{Tr}(\bmF_j^H \bmW_j \bmF_j \bmJ_{j,\theta}^T) \bmK_{j,\theta}, 
 \end{aligned}
 \end{equation}

\begin{equation}
\begin{aligned} \footnotesize
    \bmT = & \hat{\bmH}_{\theta,k} \widetilde{\bmU_k} \hat{\bmH}_{\theta,k}^H + \mbox{Tr}(\widetilde{\bmU_k} \bmJ_{\theta,k}^T) \bmK_{\theta,k} + \hat{\bmH}_{\theta,k} \widetilde{\bmU_k} \hat{\bmH}_{\theta,k}^H \\& + \mbox{Tr}(\widetilde{\bmU_k} \bmJ_{\theta,k}^T) \bmK_{\theta,k} + \hat{\bmH}_{\theta,0} \widetilde{\bmV_j} \hat{\bmH}_{\theta,0}^H + 
  \mbox{Tr}(\widetilde{\bmV_j} \bmJ_{\theta,0}^T) \bmK_{\theta,0}
    \\& + \hat{\bmH}_{\theta,0} \widetilde{\bmV_j}  \hat{\bmH}_{\theta,0}^H  + \mbox{Tr}(\widetilde{\bmV_j} \bmJ_{\theta,0}^T) \bmK_{\theta,0} +   \hat{\bmH}_{\theta,0} \widetilde{\bmV_j}  \hat{\bmH}_{\theta,0}^H \\& + \mbox{Tr}(\widetilde{\bmV_j} \bmJ_{\theta,0}^T) \bmK_{\theta,0} + \hat{\bmH}_{\theta,k} \widetilde{\bmU_k} \hat{\bmH}_{\theta,k}^H + \mbox{Tr}(\widetilde{\bmU_k} \bmJ_{\theta,k}^T) \bmK_{\theta,k} \\& + \hat{\bmH}_{\theta,k} \widetilde{\bmU_k} \hat{\bmH}_{\theta,k}^H + \mbox{Tr}(\widetilde{\bmU_k} \bmJ_{\theta,k}^T) \bmK_{\theta,k}.
\end{aligned}
\end{equation}
 \def\baselinestretch{0.85}

 \section*{Acknowledgement}
This work is supported by the Luxembourg National Fund (FNR)-RISOTTI–the Reconfigurable Intelligent Surfaces for Smart Cities under Project FNR/C20/IS/14773976/RISOTTI.

\bibliographystyle{IEEEtran}
\bibliography{main}

\end{document}